\documentclass[twocolumn, english, prl, superscriptaddress,floatfix, nofootinbib]{revtex4-1}

\usepackage[pdftex]{graphicx,color}
\usepackage{bm}
\usepackage{amsmath,amssymb}
\usepackage[T1]{fontenc}
\usepackage[utf8]{inputenc}
\usepackage[english]{babel}
\usepackage{lmodern}
\usepackage[dvipsnames]{xcolor}
\usepackage[breaklinks=true,colorlinks=true,
linkcolor=blue,urlcolor=Blue,citecolor=MidnightBlue,
bookmarks=true,bookmarksopenlevel=2]{hyperref}

\definecolor{valecol}{rgb}{0,0.5, 1.}
\newcommand{\vale}[1]{\textcolor{valecol}{#1}}

\definecolor{leacol}{rgb}{1.,0.5, 0}

\def \om    {\Omega}

\def \om0m {\Omega_{0\rm m}}

\newcommand{\fss}{{\rm{\it f\sigma}}_8}
\newcommand{\newc}{\newcommand}
\newc{\D}{\partial}
\newc{\ie}{{\it i.e.} }
\newc{\eg}{{\it e.g.} }
\newc{\etc}{{\it etc.} }
\newc{\etal}{{\it et al.}}
\newc{\lcdm}{$\Lambda$CDM }
\newc{\lcdmnospace}{$\Lambda$CDM}
\newc{\wcdm}{$w$CDM }
\newc{\plcdm}{Planck/$\Lambda$CDM }
\newc{\plcdmnospace}{Planck/$\Lambda$CDM}
\newc{\wlcdm}{WMAP7/$\Lambda$CDM }
\newc{\wlcdmnospace}{WMAP7/$\Lambda$CDM}

\newc{\ra}{\Rightarrow}
\newc{\fs}{$f\sigma_8$}
\newc{\fsz}{$f\sigma_8(z)$}

\newc{\bea}{\begin{eqnarray*}}
\newc{\eea}{\end{eqnarray*}}
\newc{\be}{\begin{equation}}
\newc{\ee}{\end{equation}}
\newc{\ba}{\begin{eqnarray}}
\newc{\ea}{\end{eqnarray}}

\begin{document}

\title{A rapid transition of \boldmath{$G_{\rm eff}$} at \boldmath{$z_t \simeq 0.01$} as a possible solution of the Hubble and growth tensions}

\author{Valerio Marra}\email{valerio.marra@me.com}
\affiliation{Núcleo de Astrofísica e Cosmologia \& Departamento de Física, Universidade Federal do Espírito Santo, 29075-910, Vitória, ES, Brazil}
\affiliation{INAF -- Osservatorio Astronomico di Trieste, via Tiepolo 11, 34131, Trieste, Italy}
\affiliation{IFPU -- Institute for Fundamental Physics of the Universe, via Beirut 2, 34151, Trieste, Italy}
\author{Leandros Perivolaropoulos}\email{leandros@uoi.gr}
\affiliation{Department of Physics, University of Ioannina, GR-45110, Ioannina, Greece}


\begin{abstract}
The mismatch in the value of the Hubble constant from low- and high-redshift observations may be recast as a discrepancy between the low- and high-redshift determinations of the luminosity of Type Ia supernovae, the latter featuring an absolute magnitude which is $\approx 0.2$~mag lower.
Here, we propose that a rapid transition in the value of the relative effective gravitational constant $\mu_G\equiv\frac{G_{\rm eff}}{G_N}$ at $z_t\simeq 0.01$ could explain the lower luminosity (higher magnitude) of local supernovae, thus solving the $H_0$ crisis.
In other words,
here the tension is solved by featuring a transition at the perturbative rather than background level.
A model that features $\mu_G = 1$ for $z \lesssim 0.01$ but $\mu_G \simeq 0.9$ for $z \gtrsim 0.01$ is trivially consistent with local gravitational constraints but would raise the Chandrasekhar mass and so decrease the absolute magnitude of Type Ia supernovae at $z \gtrsim 0.01$ by the required value of $\approx 0.2$~mag.
Such a rapid transition of the effective gravitational constant would not only resolve the Hubble tension but it would also help resolve the growth tension as it would reduce the growth of density perturbations without affecting the \plcdm background expansion.
\end{abstract}
\maketitle
{\em \textbf{Introduction.}}---The mismatch in the determination of the Hubble constant using two different well-understood calibrators -- the Cepheid-calibrated supernovae Ia (SnIa) and the CMB-calibrated sound horizon at last scattering -- is perhaps the most important open problem of modern cosmology~\cite{Knox:2019rjx,DiValentino:2021izs,Perivolaropoulos:2021jda}.
The latest CMB constraint from the Planck Collaboration \cite[][table 2]{Aghanim:2018eyx} is:
\begin{align} \label{P18}
H^{\rm P18}_0 = 67.36 \pm 0.54 \text{ km s}^{-1} {\rm Mpc}^{-1} \,,
\end{align}
while the latest local determination by SH0ES reads~\cite{Riess:2020fzl}:
\begin{align} \label{R20}
H^{\rm R20}_0 = 73.2 \pm 1.3 \text{ km s}^{-1} {\rm Mpc}^{-1} \,.
\end{align}
This 9\% discrepancy, corresponding to more than 4$\sigma$, is the so-called $H_0$ crisis.

{\em \textbf{Mismatch in the SnIa absolute magnitude.}}---As pointed out in \cite{Camarena:2021jlr,Efstathiou:2021ocp}, it is useful to look at the $H_0$ crisis via the corresponding constraints on the absolute magnitude $M_B$ of SnIa.
Propagating, via BAO measurements, the CMB constraint on the recombination sound horizon $r_d$ to the SnIa absolute magnitude $M_B$ using the parametric-free inverse distance ladder of \cite{Camarena:2019rmj}, one obtains:
\begin{align} \label{P18M}
M_B^{\rm P18} = - 19.401 \pm 0.027 \text{ mag} \,,
\end{align}
while, using the demarginalization method of \cite{Camarena:2019moy}, the constraint of equation~\eqref{R20} implies:
\begin{align} \label{R20M}
M_B^{\rm R20} = -19.244 \pm 0.037 \text{ mag} \,.
\end{align}
These two determinations are in tension at $3.4\sigma$.
Note that these values are relative to the calibration of the Pantheon supernova dataset~\cite{Scolnic:2017caz}:
supernovae Ia become standard candles only after standardization and the method used to fit supernova Ia light curves, and its parameters, can influence the inferred value of $M_B$.

The local constraint of equation~\eqref{R20M} comes from the astrophysical properties of anchors, Cepheids and the white dwarfs responsible for the SnIa explosions. The CMB determination of equation~\eqref{P18M} comes instead from the combined constraining power of CMB, BAO and SnIa observations. 
More precisely, first, CMB and BAO constrain the luminosity distance-redshift relation $d_L(z)$ and so the distance modulus $\mu(z)$:
\begin{align} \label{muu}
\mu(z) \equiv m_B(z) - M_B = 5\log_{10}\left[\frac{d_L(z)}{10 \text{pc}} \right]   \,.
\end{align}
Then, the Pantheon dataset, by constraining the SnIa apparent magnitudes $m_B$, produces a calibration  of~$M_B$.
Here, one exploits the fact that SnIa are standard candles with an \textit{a priori} unknown $M_B$.

Seen from this point of view, the $H_0$ crisis is more of an astrophysical problem than a cosmological one.
Indeed, as the local calibration is performed via Cepheid stars at redshifts less than $0.01$, it seems that a solution to this crisis could be the existence of a mechanism that changes the physics of SnIa so that they are dimmer at $z<0.01$ as compared to their luminosity at higher $z$.

\begin{figure}
\centering
\includegraphics[width = \columnwidth]{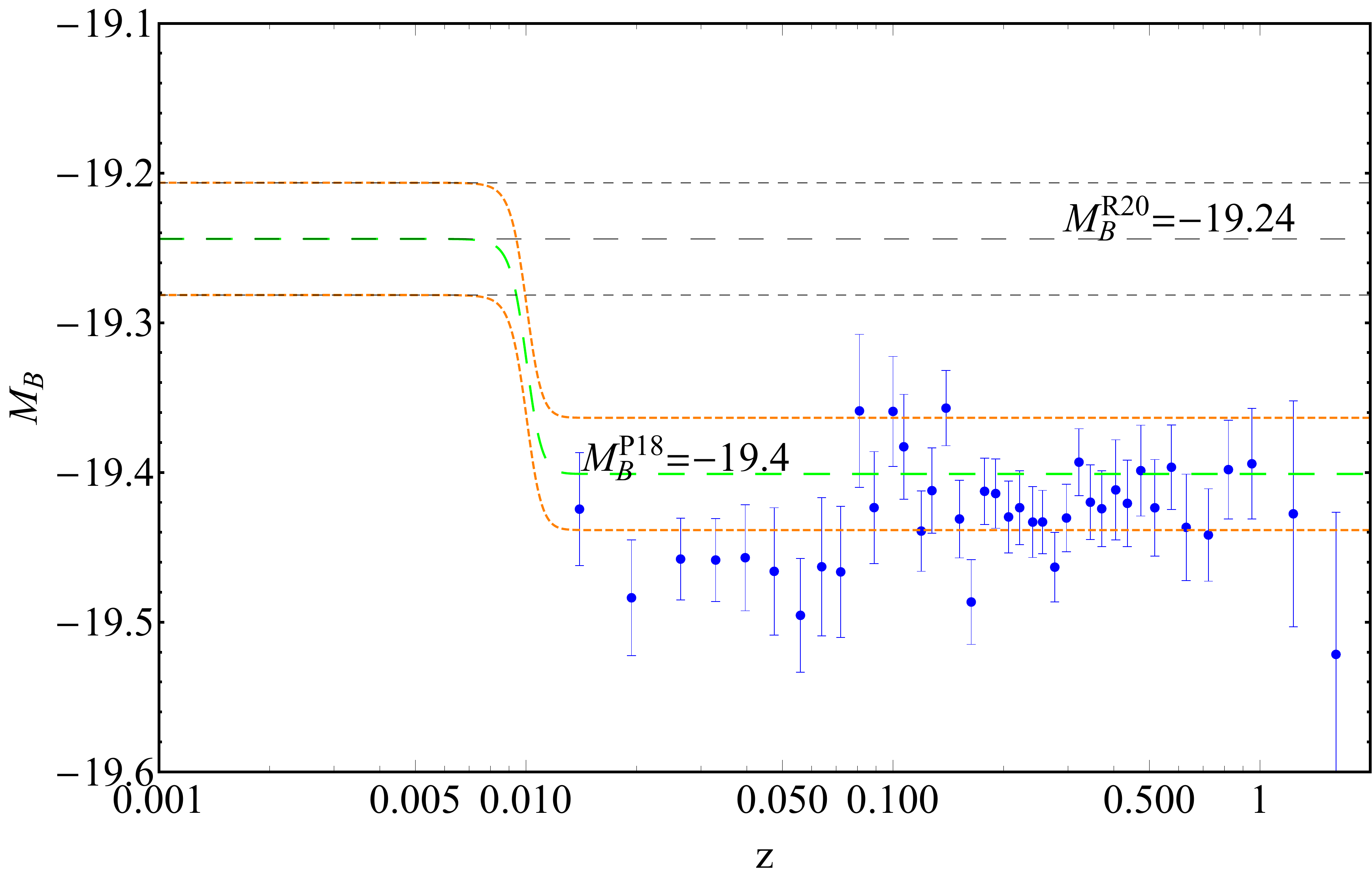}
\caption{
The inferred SnIa absolute magnitudes $M_{B,i}= m_{B,i}-\mu(z_i)$ of the binned Pantheon sample~\cite{Scolnic:2017caz} under the assumption of a \plcdm luminosity distance.
The data are inconsistent with the $M_B^{\rm R20}$ of eq.~\eqref{R20M} but they become consistent if there is a transition in the absolute magnitude with amplitude $\Delta M_B \approx-0.2$. Note that in this model there is no transition in the Hubble function $H(z)$.}
\label{fig:mbtrans}
\end{figure}

A simple phenomenological approach is to assume a transition in the value of the absolute magnitude $M_B$ at a redshift $z_t\simeq 0.01$~\cite{Alestas:2020zol}:
\begin{equation}
M_B(z) =\left\lbrace \!\!
\renewcommand{\arraystretch}{1.5}
\begin{array}{lll}
M_B^{\rm R20}  & &\text{ if } z\le z_t  \\
M_B^{\rm R20} + \Delta M_B & &\text{ if }  z> z_t 
\end{array}
\right. \,,
\label{eq:MBt}
\end{equation} 
where the needed gap in luminosity is approximately related to the corresponding gap in  $H_0$ according to (see equation~\eqref{muu}):
\begin{align} \label{rel}
\Delta M_B \equiv  M_B^{\rm P18} - M_B^{\rm R20} \approx 5 \log_{10} \frac{H^{\rm P18}_0}{H^{\rm R20}_0} \approx - 0.2  \,.
\end{align}
Figure~\ref{fig:mbtrans} illustrates this scenario showing the inferred SnIa absolute magnitudes $M_{B,i}= m_{B,i}-\mu(z_i)$ of the binned Pantheon sample~\cite{Scolnic:2017caz} for a \plcdm luminosity distance.
The data are inconsistent with the determination of equation~\eqref{R20M} but they become consistent if there is the $M_B$ transition  proposed in equation~\eqref{eq:MBt}.
In other words, a model which features this transition and with a phenomenology similar to the one of the standard $\Lambda$CDM model  would be compatible with the local Cepheid calibration, SnIa, BAO and CMB.
Note that here, contrary to the models discussed in~\cite{Mortonson:2009qq,Benevento:2020fev,Dhawan:2020xmp,Alestas:2020zol,Camarena:2021jlr,Efstathiou:2021ocp}, there is no transition in the Hubble function $H(z)$, which remains as in the standard $\Lambda$CDM model. As discussed in the introduction, this model addresses the $H_0$ tension directly at the level of the supernova absolute luminosity.
In other words, differently from previous works,
here the tension is solved by featuring a transition at the perturbative rather than background level.

{\em \textbf{The gravitational transition model and the \boldmath{$H_0$} crisis.}}---%
As we will now argue, a sudden change in the value of the effective gravitational  constant $G_{\rm eff}$ could induce the $M_B$ transition of equation~\eqref{eq:MBt} and explain, in addition,
the reduced growth rate of perturbations that is suggested by the  $\sigma_8-\Omega_{0m}$ tension.

By ``effective gravitational constant'' we refer to the strength of the gravitational interaction and not to the Planck mass related $G_*$ which determines the cosmological expansion rate~\cite{EspositoFarese:2000ij} (see \cite{Hanimeli:2019wrt} for a model with a  time-dependent $G_*$ that explains the accelerated expansion without a cosmological constant).
In other words, we are considering a scenario in which perturbations are suppressed but the background expansion does not deviate from the one of the phenomenologically successful standard $\Lambda$CDM model, for which the low-redshift expansion rate $H(z)$ is of the form
\be 
H(z) =H_0^{\rm P18} \sqrt{\Omega_{0m}(1+z)^3 +1-\Omega_{0m}} \,.
\label{eq:hzwcdm} 
\ee 

A SnIa explosion occurs when the mass of a white dwarf reaches the critical Chandrasekhar mass $m_c$ by accreting matter from a companion.  The constancy in time of this critical mass is the cornerstone on which the `SnIa standard candle' hypothesis is based. Even though  SnIa at $z>0.01$ seem consistent with the hypothesis of a constant $M_B$~\cite{Sapone:2020wwz} and, therefore, a smooth variation of $m_c$ is unlikely, a sudden transition of $m_c$ at $z<0.01$ can not be easily excluded.

The Chandrasekhar mass depends on both $G_{\rm eff}$ and the mass per electron $m^{\prime}$ according to~\cite{Amendola:1999vu}:
\begin{equation}
m_{c}\simeq \frac{3}{m^{\prime }}\left( \frac{\hbar c}{G_{\rm eff}}\right) ^{3/2} \,.
\end{equation}
In the standard model, this result is  independent of the accretion history and of the white dwarf progenitor details~\cite{Woosley:1986ta} and leads to $m_c\simeq 1.4 M_\odot$. A possible sudden transition of a fundamental constant would induce a corresponding transition of both the Chandrasekhar mass and therefore the SnIa peak absolute luminosity $L$~\cite{Arnett:1982ioj}. A reasonable hypothesis is that $L$ increases as $m_c$ increases or equivalently as $G_{\rm eff}$ (and/or $m^\prime$) decreases~\cite{Gaztanaga:2001fh}. 

Assuming a fixed $m^{\prime}$ (and fine structure constant), this may be expressed via a power law dependence
\begin{equation} \label{alla}
L\sim G_{\rm eff}^{-\alpha } \,,
\end{equation}
where $\alpha >0$ is of $O(1)$. The simplest hypothesis $L\sim m_c$ leads to  $\alpha=3/2$. This value of $\alpha$ will be assumed in the present analysis~\cite{Gaztanaga:2001fh}.
Since the SnIa absolute magnitude is connected with the absolute luminosity via 
\be
M_B-M_{0B}= -\frac{5}{2}\log_{10} \frac{L}{L_{0}} \,,
\label{lmconn}
\ee
where the index $_0$ indicates the local (present) values, it is clear that a decrease of $G_{\rm eff}$ would lead to an increase of $L$ and a decrease of $M_B$.

It then follows that the $M_B$ transition of equation~\eqref{eq:MBt}, and so the $H_0$ crisis, could be explained by the following evolution of the effective gravitational constant:
\begin{equation}
\mu_G(z)\equiv \frac{G_{\rm eff}}{G_{\rm N}} =\left\lbrace \!\!
\renewcommand{\arraystretch}{1.5}
\begin{array}{lll}
1 \equiv \mu_G^<   & &\text{ if } z\le z_t  \\ 
1+ \Delta \mu_G \equiv \mu_G^> & &\text{ if }  z> z_t  
\end{array}
\right. \,,
\label{eq:mut}
\end{equation} 
where we considered the perturbation parameter $\mu_G$~(see, e.g., \cite{Aghanim:2018eyx}), $G_{\rm N}$ is the locally measured Newton's constant, and the change in $\mu_G$ is connected to the change in the SnIa absolute magnitude via:
\begin{align}
\Delta M_B = - \frac{5}{2} \log_{10} \frac{L^{\rm P18}}{L^{\rm R20}}
=  \frac{15}{4} \log_{10} \mu_G^> \,,
\end{align}
where $L^{\rm P18}$ and $L^{\rm R20}$ are the CMB-calibrated and Cepheid-calibrated SnIa luminosities, respectively, so that:
\begin{align} \label{demu}
\Delta \mu_G = 10^{\frac{4}{15} \Delta M_B} -1 \approx -0.12 \,. 
\end{align}

{\em \textbf{Addressing the growth tension.}}---Such a $z>z_t$ suppression of the gravitational interaction is trivially consistent with local constraints on Newton's constant and it  has the potential to explain the growth tension by decreasing the growth rate of cosmological matter fluctuations $\delta(z) \equiv \frac{\delta \rho}{\rho}(z)$ according to the dynamical linear growth equation:
\begin{equation}
\delta'' \!+\! \left[ \frac{(H^2)'}{2~H^2} \!-\!
\frac{1}{1+z}\right]\delta'
\approx  \frac{3 H_0^2}{2 H^2} (1+z) \, \mu_G(z) \, \Omega_{0m}  \, \delta ,
\label{eq:odedeltaz}
\end{equation}
where $'$ denotes derivative with respect to redshift $z$. In fact, a 12\% reduction of $\mu_G$ would fit the same data with a 12\% larger~$\Omega_{0m}$.

\begin{figure}
\centering
\includegraphics[width = 0.5 \textwidth]{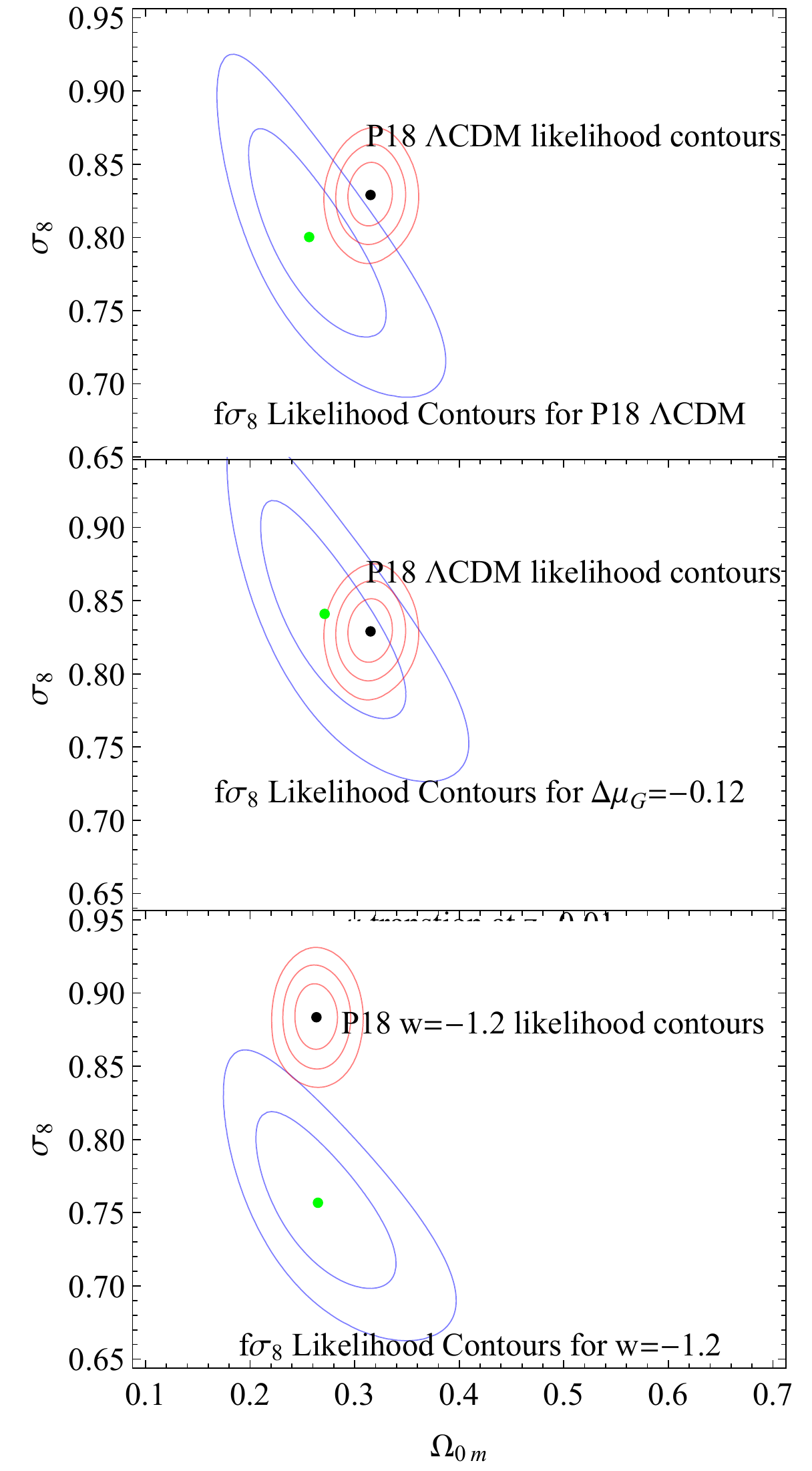}
\caption{The $\sigma_8-\Omega_{0m}$ constraints (blue) from redshift space distortion data for \plcdm (upper panel),  the $\mu$-transition model ($z_t=0.01$ and $\Delta \mu_G=-0.12$) that actually resolves the $H_0$ crisis (middle panel) and $w$CDM with $w=-1.2$ that attempts to resolve the $H_0$ tension via a smooth deformation of $H(z)$ (lower panel).}
\label{fig:grtens}
\end{figure} 

Dynamical probes of cosmological perturbations including cluster counts (CC)~\cite{Rozo:2009jj,Rapetti:2008rm,Bocquet:2014lmj,Ruiz:2014hma}, weak lensing (WL)~\cite{Schmidt:2008hc,kids1,cfhtlens,Joudaki:2017zdt,Troxel:2017xyo,kids2,des3,Abbott:2018xao} and redshift-space distortions (RSD)~\cite{Samushia:2012iq,Macaulay:2013swa,Johnson:2015aaa,Nesseris:2017vor,Kazantzidis:2018rnb} consistently favor a lower value of the matter density parameter $\Omega_{0m}$ in the context of General Relativity (GR). This implies weaker gravitational growth of perturbations than the growth indicated by GR in  the  context  of a Planck$18/\Lambda$CDM background geometry at about $2-3\sigma$ level \cite{Macaulay:2013swa,Nesseris:2017vor,Kazantzidis:2018rnb}. This weakened growth is  quantified by considering the parameter $\sigma_8$, defined as the matter density rms  fluctuations within spheres of radius $8 h^{-1}$Mpc at $z=0$.
The  value of $\sigma_8$ is connected with the amplitude of the primordial fluctuations and is determined by the growth rate of cosmological fluctuations.
A useful  bias-free statistic probed by RSD data is the quantity $f\sigma_8$:
\begin{align}
\fss(a)=\frac{\sigma_8}{\delta(a=1)}~a~\delta'(a,\Omega_{0m},\mu_G) \,,
\label{eq:fs8}
\end{align}
where $\delta$ is obtained by solving equation~(\ref{eq:odedeltaz}). This theoretical prediction may now be used to constrain, via \fs data, the parameters $\Omega_{0m}$, $\sigma_8$ and $\mu_G(z)$ in the context of a specific background $H(z)$ and a parametrization $\mu_G(z)$.
In the present analysis we assume the step-like transition of $\mu_g(z)$ proposed in equation~(\ref{eq:mut}) and either a \lcdm background or a fixed dark energy equation of state $w$.

In the context of a \lcdm background $H(z)$, RSD and WL data  are well fit by~\cite{Abbott:2017wcz,Skara:2019usd,Alestas:2020mvb}:
\begin{align}
\Omega_{0m}^{\rm growth} = (1+\Delta \mu_G)\, \Omega_{0m}^{\rm P18} =0.256^{+0.023}_{-0.031} \,,
\label{omdm}
\end{align}
which, adopting $\Omega_{0m}^{\rm P18}=0.3153\pm 0.0073$~\cite{Aghanim:2018eyx}, implies:
\begin{align}
\Delta \mu_G=-0.19\pm 0.09 \,,
\end{align}
which is in good agreement with the value required to explain the $M_B$ transition (see eq.~\eqref{demu}), and is compatible with CMB~\cite{Aghanim:2018eyx} and gravitational wave~\cite{Vijaykumar:2020nzc} constraints.

This approximate estimation of $\Delta \mu_G$ relies on the fact that in eq.~\eqref{eq:odedeltaz} the relevant parameter that drives the growth of perturbations is the product $\mu_G \times \Omega_{0m}$. The actual constraints from the full fit to the growth RSD data are presented in Figure~\ref{fig:grtens}, where
we show the  $\sigma_8-\Omega_{0m}$ likelihood contours (blue) from RSD data for \plcdm (upper panel),  the $\mu_G$-transition model ($z_t=0.01$ and $\Delta \mu_G=-0.12$) that actually resolves the $H_0$ tension (middle panel), and $w$CDM with fixed dark enrgy equation of state parameter $w=-1.2$ that attempts to resolve the $H_0$ tension via a smooth deformation of $H(z)$\cite{Alestas:2020mvb} (lower panel).
We adopted the conservative robust $f\sigma_8$ dataset of \cite[][table 2]{Sagredo:2018ahx}, which is a subset of the up-to-date compilation presented in~\cite{Skara:2019usd}. 
Superposed are the corresponding \plcdm CMB TT likelihood contours (red). The growth tension is resolved only in the case of the $\mu_G$-transition model (middle panel), while the smooth $H(z)$ deformation approach (lower panel) worsens the growth tension~\cite{Alestas:2020mvb,Alestas:2021xes} seen in the upper panel in the context of \plcdm.

{\em \textbf{Discussion.}}---We have showed that a rapid 10\% increase of the effective gravitational constant roughly \vale{100--150} million years ago ($z_t \simeq 0.01$)
can solve simultaneously the $H_0$ crisis and the $\sigma_8-\Omega_{0m}$ growth tension. A physical model where such a transition could be realized is a scalar-tensor theory with a step-like scalar field potential $V(\phi)$ and/or with an abrupt feature in the functional form of the nominimal coupling function $F(\phi)$~\cite{EspositoFarese:2000ij,Ashoorioon:2014yua,Dainotti:2021pqg}. In this theory the gravitational interactions are determined by  $G_{\rm eff}=\frac{1}{8\pi F}\frac{2 F+4F_{,\phi}^2}{2 F+3F_{,\phi}^2}$ while the background expansion is controlled by the Planck mass which corresponds to $G_*=\frac{1}{8\pi F}$. Alternatively, a quantum tunelling between two distinct degenerate vacua of a scalar tensor potential $V(\phi)$ would also induce such a gravitational phenomenological transition for an appropriate form of a nonminimal coupling $F(\phi)$.

One could test the $\mu_G$ transition model on cosmological and astrophysical scales.
First, future surveys will tightly constrain the growth of perturbation at $0<z<2$~\cite{Marra:2019lyc}, producing constraints on $\mu_G$ that may rule out the  model.

Second, low-redshift gravitational wave observations could not only rule out the mechanism here proposed  but, if detected, map it through redshift. For example, gravitational waves of merging binary neutron stars carry information about the value of $G_{\rm eff}$ at the time the merger took place. Thus bounds can be placed on the variation of $G_{\rm eff}$ between the merger time and the present time~\cite{Vijaykumar:2020nzc} assuming that the actual masses of the neutron stars are consistent with the theoretically allowed range.
Similarly, one could constrain $G_{\rm eff}$ via the gap in the mass spectrum of black holes due to the existence of pair-instability supernovae~\cite{Ezquiaga:2020tns}.
Alternatively, one may use standard sirens to measure the low-$z$ luminosity distance and compare it with the corresponding luminosity distance obtained with SnIa standard candles where the Chandrasekhar mass and $G_{\rm eff}$ are assumed  constant.
Future standard siren gravitational wave measurements are expected to constrain (or detect) such variations of $G_{\rm eff}$ at a level of $1.5\%$~\cite{Zhao:2018gwk}.
Lastly, we would like to point out that the detection of the electromagnetic counterpart GRB170817A to the gravitational wave signal GW170817 \cite{GBM:2017lvd} has led to important constraints  on the propagation speed of GWs, which have ruled out a wide range of Horndeski's modified gravity theories that predict evolving $G_{\rm eff}$. However, a wide range of scalar tensor theories, as the ones previously discussed, which also predict the cosmological evolution of $G_{\rm eff}$ for proper forms of the non-minimal coupling and the scalar field potential, have remained viable and consistent with these constraints \cite{Baker:2017hug,Creminelli:2017sry,Sakstein:2017xjx,Ezquiaga:2017ekz}.

The $\mu_G$ transition scenario may also be constrained via stellar constraints. Indeed, one expects the evolution of a star to be strongly dependent on the strength of the gravitational interaction.
A variation of $G_{\rm eff}$ would indeed affect the hydrostatic equilibrium of the star and in particular its pressure profile and temperature. 
This, in turn, will then affect the nuclear reaction rates so that the lifetimes of the various stars will be modified~\cite{Uzan:2010pm}.
From this point of view, it is interesting that an analysis of recent Tully-Fisher data~\cite{Alestas:2021nmi} has identified hints for a transition of magnitude and sign as required by the proposed mechanism albeit at somewhat more recent times: about 70 million years ago, corresponding to $z_t \simeq 0.005$.
Also interesting is the fact that a recent analysis, which analyzes the luminosity of Cepheids in anchor galaxies and SnIa host galaxies, finds a systematic brightening of Cepheids at distances larger than about 20 Mpc~\cite[][fig.~4]{Mortsell:2021nzg}. This brightening would be enough to resolve the Hubble tension and the authors attribute it to variation of dust properties. This effect appears to be consistent with our approach and, in the context of the proposed mechanism, it would occur due to a fundamental physics transition about 70 million years ago.

Also earth paleontology data  may lead to interesting constraints on the proposed mechanism. A crude approach to this problem, discussed in~\cite{Uzan:2010pm} and originally proposed in~\cite{PhysRev.73.801}, indicates that the temperature on Earth varies according to $G^{2.25}$, assuming that the mass of the Sun remains constant and ignoring atmospheric and other climatological effects (green-house effects etc). Within this approximation, if $G$ had been smaller by $10\%$ 100--150 million years ago, the temperature of the Earth would have been about $20\%$ lower. Transitions in the temperature of the Earth of this magnitude are known to have taken place during the last 500 million years, and are attributed to climatological effects or to the meteorite that is believed to have led to the extinction of various species, including the dinosaurs, about 70 million years ago. Therefore, if atmospheric effect uncertainties are taken into account, a $10\%$ gravitational transition can not be easily {\it a priori} excluded.

Finally, an important assumption made in our analysis is that the SnIa luminosity increases with the Chandrasekhar mass.
This is intuitively correct and consistent with a large part of the literature~\cite{Amendola:1999vu,Gaztanaga:2001fh,Nesseris:2006jc,Kazantzidis:2019dvk}. However, a recent analysis~\cite{Wright:2017rsu}, using a semi-analytical model to calculate SnIa light curves in non-standard gravity has claimed that the standardized peak luminosity, the quantity that is important for supernova cosmology, may actually decrease with the Chandrasekhar mass, that is, the $\alpha$ of eq.~\eqref{alla} may have a different value.
If this was the case then the proposed model would require stronger gravity to resolve the Hubble crisis and would thus be unable to resolve simultaneously the growth tension.
Then, as discussed in, e.g., \cite{Livio:2018rue,Jha:2019svc,Flors:2019hdn}, SnIa progenitors are not necessarily Chandrasekhar-mass white dwarfs and a significant fraction can arise from sub-Chandrasekhar explosions.
While this surely calls for a more detailed analysis of the dependence of the SnIa luminosity on $G_{\rm eff}$, one must note
that the Chandrasekhar mass scale is a fundamental reference scale that plays an important role in all SnIa explosions.
However, a nontrivial relation between progenitors and SnIa could again imply that the relation of eq.~\eqref{alla} could feature a different value of $\alpha$.
Lastly, it is important to study the covariance between the supernova calibrator and Hubble-flow supernovae within the proposed model, following the methodology introduced in \cite{Dhawan:2020xmp}.
The clarification of these important issues is therefore an interesting extension of the present analysis.

\section*{Acknowledgements}

It is a pleasure to thank George Efstathiou and Sunny Vagnozzi for useful comments and discussions.
VM thanks CNPq and FAPES for partial financial support.
This project has received funding from the European Union’s Horizon 2020 research and innovation programme under the Marie Skłodowska-Curie grant agreement No 888258.
LP's  research is cofinanced by Greece and the European Union (European Social Fund - ESF) through the Operational Programme ``Human Resources Development, Education and Lifelong Learning 2014-2020'' in the context of the project
``Scalar fields in Curved Spacetimes: Soliton Solutions,
Observational Results and Gravitational Waves'' (MIS 5047648).

\raggedleft
\bibliography{Bibliography}

\end{document}